\documentclass[aps,prxquantum,twocolumn,superscriptaddress,longbibliography]{revtex4-2}

\usepackage{graphicx}    
\usepackage{physics}
\usepackage{amsmath,amssymb,bm} 
\usepackage[colorlinks=true,citecolor=blue,urlcolor=blue,linkcolor=blue]{hyperref}

\begin{document}

\title{On-Chip Levitated Neon Particle Arrays for Robust and Scalable Electron Qubits}
\author{Sosuke Inui}
\thanks{These authors contributed equally to this work.}
\affiliation{National High Magnetic Field Laboratory, 1800 East Paul Dirac Drive, Tallahassee, Florida 32310, USA}
\affiliation{Department of Mechanical and Aerospace Engineering, FAMU-FSU College of Engineering, Florida State University, Tallahassee, Florida 32310, USA}

\author{Yinghe Qi}
\thanks{These authors contributed equally to this work.}
\affiliation{National High Magnetic Field Laboratory, 1800 East Paul Dirac Drive, Tallahassee, Florida 32310, USA}
\affiliation{Department of Mechanical and Aerospace Engineering, FAMU-FSU College of Engineering, Florida State University, Tallahassee, Florida 32310, USA}

\author{Yiming Xing}
\thanks{These authors contributed equally to this work.}
\affiliation{National High Magnetic Field Laboratory, 1800 East Paul Dirac Drive, Tallahassee, Florida 32310, USA}
\affiliation{Department of Mechanical and Aerospace Engineering, FAMU-FSU College of Engineering, Florida State University, Tallahassee, Florida 32310, USA}

\author{Charles Peretti}
\affiliation{National High Magnetic Field Laboratory, 1800 East Paul Dirac Drive, Tallahassee, Florida 32310, USA}
\affiliation{Department of Mechanical and Aerospace Engineering, FAMU-FSU College of Engineering, Florida State University, Tallahassee, Florida 32310, USA}

\author{Dafei Jin}
\affiliation{Department of Physics and Astronomy, University of Notre Dame, Notre Dame, Indiana 46556, USA}

\author{Wei Guo}
\email[Corresponding: ]{wguo@magnet.fsu.edu}
\affiliation{National High Magnetic Field Laboratory, 1800 East Paul Dirac Drive, Tallahassee, Florida 32310, USA}
\affiliation{Department of Mechanical and Aerospace Engineering, FAMU-FSU College of Engineering, Florida State University, Tallahassee, Florida 32310, USA}

\date{\today}
\begin{abstract}
Electron-on-neon (eNe) qubits have recently emerged as a compelling platform for quantum computing, which combines the vacuum isolation advantages of trapped-ion qubits with the scalability of superconducting circuits. In this system, electrons are trapped in vacuum above a solid neon film deposited on superconducting microwave resonators, where they exhibit strong coupling to the resonators, coherence times of $\sim0.1$ ms, and single-qubit gate fidelities exceeding 99.97\%. A central challenge, however, is the spontaneous binding of electrons to neon surface bumps. These bumps, originating from substrate roughness, vary in size: electrons on bumps of suitable sizes within the resonator can couple to microwave photons and function as qubits, whereas those on unfavorable bumps remain inactive yet contribute to background charge noise. Moreover, both the bump landscape and the sites where electrons bind differ from run to run, leading to variable qubit characteristics that hinder scalability. To address this challenging issue, we present an on-chip magnetic-levitation architecture in which arrays of solid-neon microparticles are suspended above the processor chip to act as electron carriers. This design eliminates substrate effects while retaining strong qubit–resonator coupling and supporting inter-qubit connectivity. Our analysis further shows that the qubit transition frequency can be tuned across the gigahertz range and its anharmonicity can reach $\sim0.8$~GHz by tuning the resonator bias voltage. Together, these features suggest a promising pathway toward robust, reproducible, and scalable eNe-based quantum computing.

\end{abstract}
\pacs{xxxx}
\maketitle

\section{\label{Sec:I}Introduction}
Quantum computing has made remarkable progress across various qubit platforms \cite{Ladd2010,Zwanenburg2013,Atatre2018,Dobrovitski2013,Goldner2015,Gali2019,Zhong2019,DeLeon2021}, yet achieving large-scale, fault-tolerant operation remains a major challenge \cite{Markov2014,Singh2021,DeLeon2021}. Among the leading platforms, superconducting qubits and trapped-ion qubits have each demonstrated substantial success while also facing inherent trade-offs. Superconducting qubits offer fast on-chip microwave control and compatibility with lithographic fabrication, making them promising for large-scale integration~\cite{Wallraff2004,Blais2021,Nakamura1999,Schoelkopf2008,Clarke2008,Arute2019,Kjaergaard2020,Krantz2019,Wendin2017,Gambetta2017}. However, their coherence is often limited by material defects such as two-level systems, which induce energy relaxation and dephasing that constrain gate fidelities \cite{siddiqi2021engineering,wang2022towards}. As a result, encoding a single logical qubit requires a large number of physical qubits, creating significant hardware overhead. In contrast, trapped-ion qubits exhibit long coherence times and high gate fidelities due to their isolation in ultra-high vacuum, resulting in much lower error-correction overhead \cite{Monroe1995,Kielpinski2002,Leibfried2003,Pino2021,Bruzewicz2019}. Yet scaling ion traps remains challenging due to increasingly complex architectures, stringent multi-ion control requirements, and challenges in addressing individual ions as motional modes crowd \cite{Bruzewicz2019,Brown2021}. The complementary strengths and limitations of these systems have motivated the development of hybrid architectures that combine vacuum isolation with fast microwave control and scalability enabled by semiconductor fabrication. Such designs may ultimately unlock a transformative pathway toward fault-tolerant quantum computing.

\begin{figure}[t]
\centering
\includegraphics[width=1\linewidth]{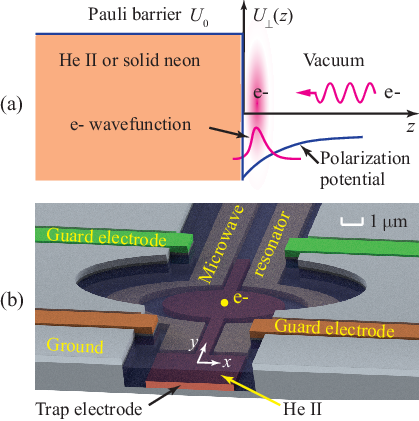}
\caption{(a) Schematic showing the Pauli barrier and surface-bound state of an electron on He~II or solid neon. (b) Schematic of a typical He~II–filled microchannel device for trapping and controlling a surface electron~\cite{Koolstra2019}.}
\label{Fig1}
\end{figure}

Building on this motivation, extensive efforts have focused on developing charge qubits based on single electrons trapped on the surface of superfluid $^4$He (He~II) \cite{Platzman1999,Schuster2010,Koolstra2019,Kawakami2019,Wang2025,Beysengulov2024,Guo2025}. As shown schematically in Fig.~\ref{Fig1}(a), when an electron approaches the He~II surface, it polarizes the helium atoms, inducing positive surface charges that create an attractive potential $U_{\bot}(z)=-\frac{\varepsilon-\varepsilon_0}{4(\varepsilon+\varepsilon_0)}\frac{e^2}{4\pi \varepsilon_0 z}$, where $\varepsilon_0$ is the vacuum permittivity, $\varepsilon = 1.05\varepsilon_0$ is the dielectric constant of He~II, and $z$ is the distance from the surface. Meanwhile, Pauli exclusion between the excess electron and $^4$He shell electrons produces an energy barrier of $\sim1$~eV, preventing the electron from entering the liquid. These combined effects confine the electron vertically, yielding a ground-state energy of about $-7.8$~K at a mean height $\langle z \rangle_0 \simeq 10$~nm above the surface \cite{Cole1970,Cole1971,Cole1969,Jin2020}. Using microchannels and patterned electrodes, the lateral motion of these surface electrons can also be confined (see Fig.~\ref{Fig1}(b)), allowing their vertical bound states, lateral motional states, or spin states to serve as qubit levels \cite{Schuster2010,Koolstra2019}. The electrons reside in high vacuum above ultraclean He~II, and their lateral and spin states can be engineered to have GHz transition energies, enabling on-chip microwave control via coplanar waveguide (CPW) resonators \cite{Schuster2010}. Substantial progress has been made on single-electron detection and manipulation \cite{Castoria2025,Beysengulov2024,koolstra2025strongcouplingmicrowavephoton,Denis2025}. However, surface fluctuations of He~II in the microchannel can cause qubit transition-line broadening and decoherence \cite{Koolstra2019}, a key challenge under active investigation.

Besides He~II, electrons can also bind to the surfaces of other quantum fluids and solids via the same mechanism \cite{Guo2025}. In particular, recent experiments implemented devices similar to that in Fig.~\ref{Fig1}(b) but with a solid-neon (SNe) film ($\sim10$~nm thick) deposited on the surface instead of He~II~\cite{Zhou2022,Zhou2024}. Owing to neon’s larger dielectric constant $\varepsilon=1.24\varepsilon_0$, the electron binds more strongly to the surface, forming a vertical ground state with energy of $-183$~K and mean height $\langle z \rangle_0 \simeq 1$~nm \cite{Cole1969,Cole1971,Cole1970,Jin2020}. With a suitable positive bias on the trap electrode, the electron’s $y$ motion remains in the ground state, while its two lowest $x$-motional states can be tuned to a transition frequency of $\sim6.2$~GHz, resonant with the microwave resonator. Since SNe has negligible surface fluctuations compared with He~II, strong electron–resonator coupling was clearly observed \cite{Zhou2024}, with a coupling strength $g/2\pi>4$~MHz and a transition linewidth $\gamma/2\pi \simeq 0.1$~MHz. Furthermore, coherence times $T_1$ and $T_2$ of $\sim0.1$~ms were measured, along with a single-qubit gate fidelity of 99.97\%, comparable with state-of-the-art superconducting qubits. Theoretically, coherence times exceeding 10~ms for the lateral motional states and about 81~s for the spin states are expected \cite{Guo2025,chen2022electron}, highlighting the exceptional promise of the electron-on-neon (eNe) platform for scalable quantum computing.

\begin{figure*}[t]
\centering
\includegraphics[width=1\linewidth]{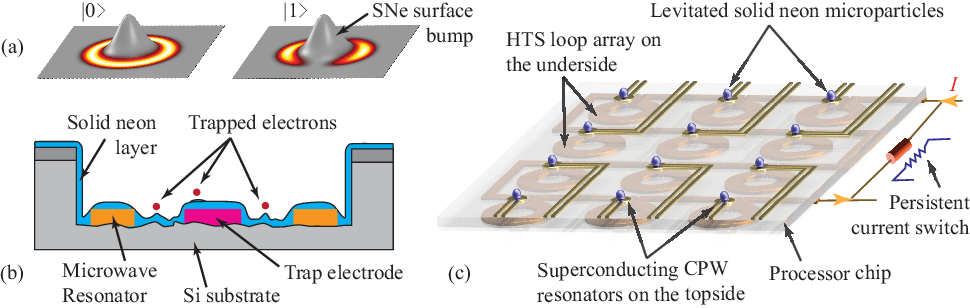}
\caption{(a) Schematic wavefunction profiles of an electron self-bound to a SNe surface bump in the ground state $\ket{0}$ and first excited state $\ket{1}$~\cite{Kanai2024}. (b) Cross-sectional schematic of the device in Fig.~\ref{Fig1}(b), showing electrons bound to surface bumps on the SNe film deposited on the device. (c) Conceptual diagram of the proposed on-chip architecture using arrays of magnetically levitated SNe microparticles as electron qubit carriers.}
\label{Fig2}
\end{figure*}

Nonetheless, a puzzling observation emerged: electrons can remain anchored in the resonator even without an applied trapping potential, pointing to an intrinsic surface-induced trapping mechanism. Our recent theoretical study revealed that electrons can spontaneously bind to bumps on the neon surface due to lateral forces arising from induced charge on the curved surface \cite{Kanai2024}, forming ring-shaped quantum states around the bump base (see Fig.~\ref{Fig2}(a)). When the bump width is suitable (i.e., $\sim$30 nm), the transition energy between the two lowest states matches the resonator photon energy, enabling qubit operations. This picture was supported experimentally \cite{Zheng2025}, which found that strongly coupled electron states appear only in devices fabricated on rough etched-silicon substrates, where the SNe surface inherits nanoscale bumps. Upon thermionic injection, electrons bind to these bumps (see Fig.~\ref{Fig2}(b)), but only those on bumps of the right size within the resonator act as qubits. Electrons trapped on bumps of unfavorable sizes remain inactive yet contribute to background charge noise, degrading coherence. Moreover, the bump landscape and binding sites vary between devices and cooldowns, leading to variable qubit characteristics. This stochastic trapping has thus emerged as a key challenge for realizing reproducible and scalable eNe qubit arrays~\cite{matkovic2025,Murch-2025}. Overcoming it requires decoupling electron confinement from substrate-induced surface morphology.

In this paper, we present an on-chip magnetic levitation architecture, shown in Fig.~\ref{Fig2}(c), where arrays of SNe microparticles are magnetically suspended above a processor chip to serve as electron carriers, thereby eliminating substrate effects. In Sec.~\ref{Sec:II}, we discuss the feasibility of levitating SNe particles using high-$T_c$ superconducting current loops patterned on the chip underside. The procedures for particle loading and electron trapping are described in Sec.~\ref{Sec:III}. Section~\ref{Sec:IV} analyzes the electron eigenstates on a neon particle under a bias voltage applied to the microwave resonator, showing that transition frequencies can be tuned across the gigahertz (GHz) range and that the qubit anharmonicity can reach up to $\sim0.8$~GHz, suitable for fast on-chip operations. In Sec.~\ref{Sec:V}, we analyze the electron–resonator coupling for single-qubit control and show how two eNe qubits can be coherently linked through resonator-mediated interactions. Collectively, these results provide a feasible solution toward reproducible and scalable eNe-based quantum processors. A brief discussion of broader applications of this platform is given in Sec.~\ref{Sec:VI}.


\section{\label{Sec:II} Magnetic levitation of Solid-neon microparticles}
When a diamagnetic material with a volume magnetic susceptibility $\chi<0$ is placed in a magnetic field $\mathbf{B}$, it experiences a potential energy density given by~\cite{Weilert1996}:
\begin{equation}
E(\mathbf{r}) = \rho g z + \frac{|\chi| B^2}{2\mu_0},
\label{Eq1}
\end{equation}
where $\rho$ is the material density, $g$ is the gravitational acceleration, and $\mu_0$ is the vacuum permeability. The condition for magnetic levitation can be obtained by setting $\partial E(\mathbf{r})/\partial z = 0$, which yields the critical field gradient $\left[\partial B^2(\mathbf{r})/\partial z\right]_c=-\mu_0 g \rho/|\chi|$.
Table~\ref{Table1} lists the critical gradients for several materials. Due to the small $|\chi|$ values of typical diamagnetic materials, achieving levitation generally requires large magnetic field gradients. In the past, levitation of He~II droplets with diameters ranging from $100~\mu\mathrm{m}$ to $1~\mathrm{cm}$ was realized using large multi-turn superconducting coils carrying currents exceeding $10^2$~A~\cite{Weilert1996, Whitaker1999, weilert1997magnetic}. For solid neon, the required gradient is even larger. Nevertheless, such high gradients can be produced on-chip with a single current loop, provided the loop size is reasonably small.

\begin{table}[t]
\caption{\label{Table1}
Comparison of magnetic levitation parameters for water at ambient conditions, He~II, and SNe.}
\begin{ruledtabular}
\begin{tabular}{lccc}
\textbf{Material} & $\rho$ (g/cm$^3$) & $\chi$ &
$|\partial B^2/\partial z|_c$ (T$^2$/cm) \\
\hline
Water & 1.00 & $-9.04\times10^{-6}$ & -13.6 \\
He~II & 0.145 & $-8.6\times10^{-7}$ & -20.7 \\
SNe & 1.44 & $-6.25\times10^{-6}$ & -28.4 \\
\end{tabular}
\end{ruledtabular}
\end{table}

\begin{figure}[t]
\centering
\includegraphics[width=1.0\linewidth]{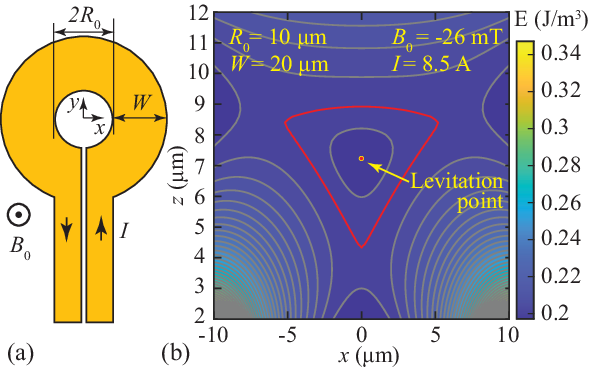}
\caption{(a) Schematic showing the HTS loop parameters used in the analysis. (b) Calculated potential energy density $E(\mathbf{r})$ for solid neon in the magnetic field generated by a current loop with the specified parameters. Contours are plotted with an energy density interval of $3.94\times10^{-3}$ J/m$^3$.}
\label{Fig3}
\end{figure}

\begin{figure*}[t]
\centering
\includegraphics[width=1.0\linewidth]{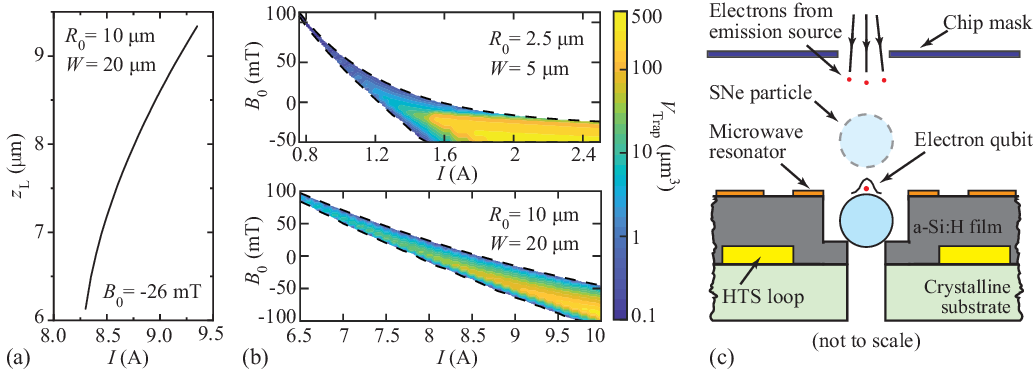}
\caption{(a) The height of the levitation point $z_L$ as a function of $I$ for the loop analyzed in Fig.~\ref{Fig3}(b). (b) Calculated trap volume $V_{\text{trap}}$ as a function of $I$ and $B_0$ for two loops of different sizes. (c) Schematic of the device near a levitation site.}
\label{Fig4}
\end{figure*}

To illustrate this, we consider an on-chip current loop with an inner radius $R_0$ and width $W$, carrying a current $I$ in the presence of a uniform background bias field $B_0$ directed along the $z$ axis, perpendicular to the loop plane, as shown in Fig.~\ref{Fig3}(a). The uniform bias field $B_0$, which can be generated by a superconducting magnet commonly used in dilution refrigerators, provides an additional degree of tunability~\cite{Sanavandi2021}. The magnetic field induced by the loop, $\mathbf{B}_L(\mathbf{r})$, in three-dimensional space above the loop can be calculated using finite-element simulations (see Appendix~\ref{App-A}). From this, we can obtain the total field $\mathbf{B}=\mathbf{B}_0+\mathbf{B}_L$ and evaluate the potential energy density $E(\mathbf{r})$ for solid neon. The result for a representative loop with $R_0=10$~$\mu$m, $W=20$~$\mu$m, $I=8.5$~A, and $B_0=-26$~mT is shown in Fig.~\ref{Fig3}(b). A trapping region appears above the loop plane, outlined by the red contour, where $E(\mathbf{r})$ decreases toward the center, i.e., the levitation point where the net force vanishes. When a solid-neon particle is placed within the trapping region, it cannot escape because of the energy barrier when moving across the trap boundary. The particle thus tends to settle at the levitation point. The height of the levitation point, $z_L$, can be conveniently adjusted by varying the loop current, as shown in Fig.~\ref{Fig4}(a), which in turn allows control of the particle’s vertical position.

Besides $z_L$, the size of the trapping region can also be tuned. To demonstrate this, we calculate the trap volume enclosed by the red contour, $V_{\text{trap}}$, as a function of $I$ and $B_0$ for two loops of different sizes, as illustrated in Fig.~\ref{Fig4}(b). The results show that $V_{\text{trap}}$ on the order of several hundred $\mu$m$^3$ can be achieved with suitable combinations of $I$ and $B_0$ for both loops. In general, loops with smaller $R_0$ require lower current $I$ to create a stable trap, whereas larger loops can achieve a much greater $V_{\text{trap}}$ when the current is sufficient. Although the trapping region is not spherical, it can readily accommodate a SNe particle with a diameter of several microns.

Depending on the exact values of $R_0$, $W$, and $B_0$, the loop current $I$ required for levitating SNe particles can range from below 1~A to over 10~A. Assuming a loop film thickness of $\delta=5~\mu$m, the corresponding current density $J = I/(W\delta)$ lies between $10^5$ and $10^7$~A/cm$^2$. Nonetheless, such high current densities are attainable using magnesium diboride (MgB$_2$) or high-$T_c$ superconducting (HTS) materials such as rare-earth barium copper oxide (REBCO)~\cite{Larbalestier2001,Foltyn1999}. Critical current densities up to $J_c = 5\times10^7$~A/cm$^2$ have been achieved in both MgB$_2$~\cite{zhang2015fabrication} and REBCO~\cite{Stangl2021}. Even at $B=3$~T and $T=30$~K, $J_c \approx 1.5\times10^7$~A/cm$^2$ has been reported~\cite{Majkic2018}, which is sufficient for levitating SNe particles at an initial temperature of $\sim25$~K (see Sec.~\ref{Sec:III}).

To fabricate the loops, one may start with an HTS film epitaxially grown on a lattice-matched substrate such as SrTiO$_3$, or as a textured thick film on sapphire or silicon~\cite{Hollmann1994,Phillips1996}. The HTS loops can then be patterned using photolithography and ion-milling (or other suitable dry-etch processes) \cite{jin1999fabrication,pearton2005dry}, followed by the deposition of a hydrogenated amorphous silicon (a-Si:H) or silicon carbide (a-SiC:H) layer of controlled thickness (see Fig.~\ref{Fig4}(c)), chosen for its exceptionally low microwave dielectric loss~\cite{buijtendorp2022hydrogenated}. Shallow wells will be fabricated at the levitation sites to accommodate the SNe particles and enhance tunability (see Sec.~\ref{Sec:III} and \ref{Sec:IV}). For qubit operation, superconducting quarter-wave ($\lambda/4$) differential CPW resonators will be patterned on the a-SiC:H layer using photolithography, metal deposition, and liftoff. Materials such as niobium titanium nitride (NbTiN) and niobium nitride (NbN) are suitable due to their strong resilience to magnetic fields. For example, NbTiN nanowire resonators have demonstrated $Q$-factors exceeding $3\times10^4$ in a 0.35-T perpendicular field~\cite{samkharadze2016high}. In our proposed device, the magnetic field at the resonator sites typically ranges from $0.1$ to $0.3$~T during particle levitation, well within the tolerance of these materials.

\section{Particle loading and electron trapping}\label{Sec:III}
\begin{figure}[t]
\centering
\includegraphics[width=1.0\linewidth]{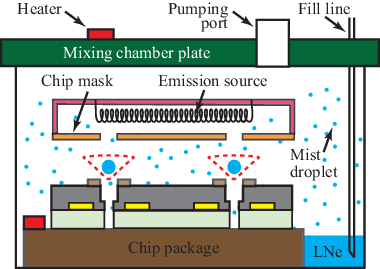}
\caption{Schematic showing the mist merging method for loading SNe particles on the chip in the experimental chamber.}
\label{Fig5}
\end{figure}

Building upon the magnetic trap design discussed above, the following procedure can be adopted for consistently loading SNe microparticles into the traps. First, the $^3$He circulation in the dilution refrigerator should be suspended, while the cryocooler remains running to keep the radiation shields cold. The experimental chamber can then be heated to about 25~K so that saturated liquid neon (LNe) can be introduced to the chamber bottom (see Fig.~\ref{Fig5}). Next, the magnetic traps can be activated by applying current to the HTS loops. Once the traps are established, the chamber may be suddenly pumped. The resulting pressure reduction causes the LNe at the chamber bottom to boil, generating a mist of LNe nanodroplets. These nanodroplets spontaneously agglomerate within the magnetic traps due to their similar $\rho/|\chi|$ ratio to that of SNe \cite{POLLACK1964}, forming larger droplets whose sizes depend on the aggregation time and are ultimately limited by the trap dimensions. During this process, the chip can be maintained at a temperature slightly above the saturation temperature of LNe to prevent condensation on its surface. This mist-agglomeration method has been successfully demonstrated previously for loading magnetically levitated He~II droplets \cite{weilert1997magnetic,Brown2023}.

For a levitated LNe droplet with a radius of a few microns, surface tension generates an internal pressure on the order of 10$^3$~Pa, far exceeding the magnetic field–induced stress variations (i.e., $\sim10^{-2}$ Pa). Consequently, the levitated droplets remain highly spherical. To convert LNe droplets into SNe particles, one may first drain the remaining LNe from the chamber bottom and then quickly evacuate the vapor. The levitated droplets cool efficiently through forced evaporation and, upon reaching the triple point of neon at 24.6 K and 0.43 bar \cite{POLLACK1964}, solidify into SNe particles. During this process, the droplet size decreases by only a few percent due to the large latent heat of LNe. The final SNe particle size, which may be tuned by controlling the evaporation duration before reaching the triple point, can be inferred from the shift in the resonator’s resonance frequency relative to the unloaded case.

\begin{figure}[t]
\centering
\includegraphics[width=1.0\linewidth]{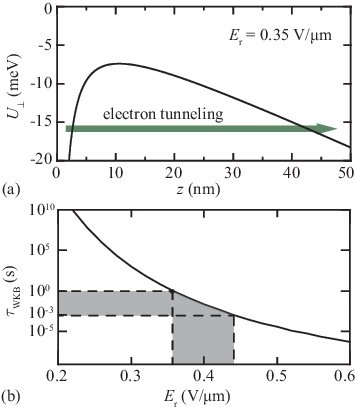}
\caption{(a) Potential energy $U_\perp$ of an electron versus distance $z$ from a SNe particle surface under an applied extraction field $E_r=0.35$~V/$\mu$m. (b) Calculated WKB tunneling lifetime $\tau_{\mathrm{WKB}}$ as a function of $E_r$. The shaded region indicates the threshold range of $E_r$ for efficient electron extraction.}
\label{Fig6}
\end{figure}

Unlike snowflakes, which form intricate branches through hexagonal lattice growth \cite{Colbeck1982}, solid neon crystallizes in an isotropic face-centered-cubic structure \cite{Niebel1997}. The uniform surface evaporation further promotes the formation of nearly spherical microparticles free from substrate-induced roughness. These levitated microparticles can oscillate within the magnetic traps. However, due to their macroscopic mass, thermal oscillations are minimal. For instance, a SNe particle with radius $R_s=3$~$\mu$m in the trap shown in Fig.~\ref{Fig3}(b) exhibits thermal motion of only a few nanometers at $100$~mK in both radial and vertical directions. Such residual oscillations can be actively suppressed using feedback-based damping techniques \cite{Bang2020}. Alternatively, one may completely suppress the oscillations by reducing the HTS loop current, allowing the SNe particles to rest in the shallow wells, as shown in Fig.~\ref{Fig4}(c).

To inject electrons, a current pulse can be applied to a tungsten-filament emission source mounted above the chip mask (Fig.~\ref{Fig5})~\cite{Zhou2022,Zhou2024}. The emitted electrons slow down as they traverse the residual neon vapor and become trapped on the SNe particle surface. To enable qubit operation with a single electron, excess electrons on an SNe particle must be removed. This can be done by applying a positive bias $V_r$ to the microwave resonator (see Sec.~\ref{Sec:IV}), which pulls the electron towards the north pole of the SNe particle and creates a perpendicular field $E_r$ at that location for electron extraction. The potential energy $U_\perp(z)$ of a surface-bound electron then develops an energy barrier through which the electron can tunnel and escape from the SNe surface. Fig.~\ref{Fig6}(a) illustrates an example potential for $E_r = 0.35$~V/$\mu$m. We have calculated the electron tunneling timescale $\tau_{\mathrm{WKB}}$ for various $E_r$ values using the WKB approximation~\cite{Cole1971} (see Appendix~\ref{App-B}). As shown in Fig.~\ref{Fig6}(b), once $E_r$ crosses a threshold region around $0.4$~V/$\mu$m, $\tau_{\mathrm{WKB}}$ drops from several seconds to below 1~ms, sufficiently short for efficient electron extraction. Such field strengths are readily attainable experimentally. The number of electrons remaining on the SNe particle can then be inferred by monitoring the resonator's transmission response as demonstrated in electron-on-LHe systems, i.e., either through changes in amplitude at a fixed frequency \cite{Koolstra2019} or through shifts in the resonance peak \cite{Yang2016}. Once it is confirmed that only a single electron remains on the surface, the chamber can be fully evacuated and the chip cooled to its base temperature for qubit operation.


\section{Quantum states and tunability of the electron qubit}\label{Sec:IV}
\begin{figure}[t]
\centering
\includegraphics[width=1.0\linewidth]{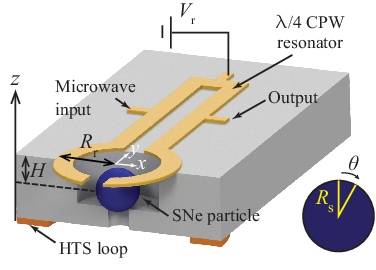}
\caption{Schematic of the $\lambda/4$ CPW resonator with a bias voltage $V_r$ applied to attract the surface-bound electron toward the north pole of the SNe particle.}
\label{Fig7}
\end{figure}

For an electron trapped on the surface of a SNe microparticle, the motion perpendicular to the surface is confined by essentially the same trapping potential $U_{\perp}$ shown in Fig.~\ref{Fig1}(a). The bound ground-state energy in this direction is about -15.8 meV, and the first excited level lies 12.7 meV above it, corresponding to a transition frequency of 3.1 THz. Under typical experimental conditions, the electron remains in its vertical ground state while being free to move laterally. In the absence of any applied electric and magnetic fields, the lateral eigenstates on an ideally spherical SNe particle are given by the spherical harmonics $Y_{l,m}(\theta,\phi)$, with eigenenergies $E_{l}=\frac{\hbar^{2}}{2m_{e}R_{s}^{2}}l(l+1)$. Since the particle radius $R_{s}$ is on the micron scale, the transition between the lowest two lateral motional states has a frequency on the order of 10 MHz, too small to resonate with the GHz microwave resonator. To address this, a positive bias voltage $V_r$ can be applied at the voltage node of the $\lambda/4$ CPW resonator, where microwave leakage is minimal~\cite{Petersson2012,Armour2013}, as shown in Fig.~\ref{Fig7}. The resulting electric field attracts the electron and localizes it near the north pole of the SNe particle. The field $E_r$ required near the north pole to raise the transition energy into the GHz range is much smaller than the threshold field for electron extraction shown in Fig.~\ref{Fig6}(b), which ensures the electron remains securely trapped near the pole. Due to this localization, moderate SNe shape deformation has little effect on the electron's lateral quantum states. In the subsequent analysis, we therefore assume an ideal spherical SNe particle.

\subsection{Electron quantum states}\label{Sec:IV-1}
To determine the lateral eigenstates of the electron, we consider the following Hamiltonian:
\begin{equation} \label{eq:Hamiltonian}
\hat{H}_{\parallel} = -\frac{\hbar^2 \nabla^2_{\theta\phi}}{2 m_e R_s^2} + \frac{eB_0}{2m_e} \hat{L}_z + \frac{e^2 B_0^2 R_s^2}{8 m_e} \sin^2\theta + U_{\parallel}(\theta),
\end{equation}
where $\nabla^2_{\theta\phi}$ is the angular part of the Laplacian operator. The second and third terms represent the orbital Zeeman and diamagnetic contributions, respectively. Since the SNe particle rests in the shallow well on the chip, a strong magnetic field is no longer required for levitation, which also relaxes the constraints on the magnetic-field resilience of the resonator materials. Nonetheless, a weak uniform field $B_0$ may still be applied to help lift unwanted degeneracies in the electron energy spectrum, as will be discussed later. The last term, $U_{\parallel}(\theta)$, denotes the electrostatic potential produced by the bias voltage $V_r$ on the resonator. In our analysis, the resonator is modeled as a charged ring of radius $R_r$, coaxial with the SNe particle and separated from its center by a distance $H$, as illustrated in Fig.~\ref{Fig7}. The distance $H$ can be adjusted either by controlling the well depth during device fabrication or by tuning the SNe particle size during the loading process. The potential $U_{\parallel}(\theta)$ depends on $R_r$, $H$, and $V_r$, and, owing to the system’s axial symmetry, can be expressed in a compact analytical form involving elliptic integrals (see Appendix~\ref{App-C} for details).

\begin{figure}[t]
\centering
\includegraphics[width=1.0\linewidth]{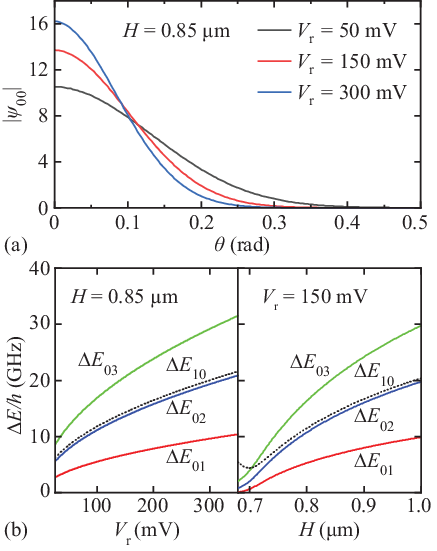}
\caption{(a) Calculated electron ground-state wavefunction profile $|\psi_{00}|$ for different bias voltages $V_r$ at a fixed distance parameter $H=0.85~\mu$m. (b) Dependence of the transition energy $\Delta E_{nm}=E_{nm}-E_{00}$ from the ground state $\psi_{00}$ to various excited states $\psi_{nm}$ on the distance parameter $H$ and bias voltage $V_r$. The results are for a representative case with $R_s=0.5$~$\mu$m, $R_r=1.5$~$\mu$m, and $B_0=-20$~mT.}
\label{Fig8}
\end{figure}

To demonstrate that an applied bias voltage can indeed increase the electron transition energy into the GHz range, we solved the Schr\"{o}dinger equation $\hat{H}_{\parallel}\ket{\psi_{n m}}=E_{n m}\ket{\psi_{n m}}$ for a representative configuration with $R_r=1.5$~$\mu$m, $R_s=0.5$~$\mu$m, and $B_0=-20$~mT, using an imaginary-time propagation method (see Appendix~\ref{App-D}). The resulting eigenfunctions $\psi_{nm}(\theta,\phi) \equiv \langle \theta,\phi | \psi_{nm} \rangle$ are characterized by two good quantum numbers: the azimuthal quantum number $m$, arising from the axial symmetry of the system, and the polar-mode quantum number $n$, which characterizes excitations along the $\theta$ direction. Fig.~\ref{Fig8}(a) shows the ground-state wavefunction profiles $|\psi_{00}(\theta)|$ obtained for different bias voltages $V_r$ at a fixed $H=0.85$~$\mu$m. The electron wavefunction is confined within a narrow polar range $\theta \lesssim 0.1$–$0.2$~rad near the north pole. As $V_r$ increases, $\psi_{00}(\theta)$ becomes progressively more localized. The calculated energy differences $\Delta E_{nm}=E_{nm}-E_{00}$ between the ground state and several excited states $\psi_{n m}$ are plotted in Fig.~\ref{Fig8}(b) as functions of $V_r$ and $H$. It is clear that the transition energy $\Delta E_{01}$ between the lowest two energy levels can be tuned continuously within the GHz range, ideal for on-chip microwave control.

\begin{figure}[t]
\centering
\includegraphics[width=1.0\linewidth]{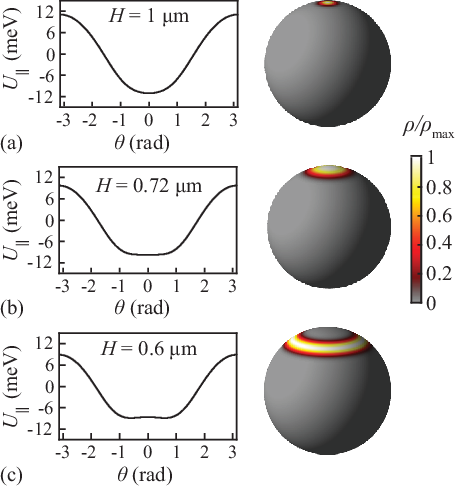}
\caption{Lateral trapping potential $U_\parallel(\theta)$ and the corresponding electron ground-state probability density $\rho = |\psi_{00}|^2$ at a fixed resonator voltage $V_r=150$~mV for distance parameters (a) $H = 1$~$\mu$m, (b) $H = 0.72$~$\mu$m, and (c) $H = 0.6$~$\mu$m.}
\label{Fig9}
\end{figure}

A few additional features of the system’s energy levels merit further discussion. First, in the absence of a magnetic field, the states $\ket{\psi_{01}}$ and $\ket{\psi_{0,-1}}$ are degenerate, which is undesirable for defining a unique qubit transition. The applied magnetic field of $B_0=-20$~mT lifts this degeneracy through the orbital Zeeman effect, introducing an energy splitting of about 0.56~GHz between the two states, sufficient for unambiguous state selectivity. The diamagnetic term in the Hamiltonian is much weaker, shifting the energies of both $\ket{\psi_{01}}$ and $\ket{\psi_{0,-1}}$ upward by only a few megahertz. Second, as shown in Fig.~\ref{Fig8}(b), the spacings between adjacent $\Delta E_{0m}$ curves are comparable, indicating that the transition energies between successive azimuthal states $\psi_{0,m}$ and $\psi_{0,m+1}$ differ only slightly. This behavior can be understood by noting that, under strong confinement near the north pole, the potential can be approximated as $U_{\parallel}(\theta) \simeq U_{\parallel}(0) + eE_r R_s (1 - \cos\theta) \simeq U_{\parallel}(0)+\tfrac{1}{2} eE_r R_s \theta^2$. In the absence of a magnetic field, the Hamiltonian reduces to that of a quantum spherical pendulum, which has been extensively studied in the context of quantum rotor systems~\cite{schmidt2015supersymmetry,PhysRevB.46.9501}. The eigenenergies in this case are approximately $E_{n m} \simeq \hbar \omega_0 (2n + |m| + 1)$, with $\omega_0 = \sqrt{e E_r / (m_e R_s)}$, leading to equal spacings between adjacent azimuthal levels. Fortunately, this degeneracy is naturally lifted by the diamagnetic contribution and by higher-order terms in the expansion of $U_{\parallel}(\theta)$ near $\theta=0$. More importantly, as will be discussed next, the spacings between azimuthal levels can be strongly tuned by varying $V_r$ and $H$.

\subsection{Qubit anharmonicity and tunability}\label{Sec:IV-2}
To quantify the nonuniformity of energy-level spacings, we can evaluate the anharmonicity parameter $\alpha = \Delta E_{02} - 2\Delta E_{01}$. As discussed earlier, when the electron is strongly confined near the north pole of the SNe particle, the energy levels are nearly equally spaced. However, a distinctive feature of our system arises from the ring-shaped resonator geometry: as the relative position between the SNe particle and the resonator changes, the lateral trapping potential $U_\parallel(\theta)$ can evolve from having a single minimum at the north pole to developing a minimum at a finite $\theta$ away from the pole. This transition occurs because $U_\parallel(\theta)$ reaches its minimum on the SNe surface approximately at the location closest to the resonator. When the center-to-center distance $H$ is large, this location coincides with the north pole; as $H$ decreases, the closest location moves to a finite $\theta$, shifting the potential minimum off-axis. Fig.~\ref{Fig9} illustrates this behavior through the calculated $U_\parallel(\theta)$ profiles for different $H$ at a fixed resonator voltage of $V_r=150$~mV. At $H = 1$~$\mu$m, $U_\parallel(\theta)$ exhibits a single minimum at the north pole and is approximately harmonic near that point; correspondingly, the ground-state wavefunction $\psi_{00}$ is nearly Gaussian, centered at the pole with a narrow spread in $\theta$. When $H$ is reduced to 0.72~$\mu$m, the potential develops a flatter bottom, and $\psi_{00}$ broadens noticeably. As $H$ decreases further to 0.6~$\mu$m, $U_\parallel(\theta)$ acquires a minimum at a finite $\theta$, and $\psi_{00}$ becomes localized along a ring around this off-pole minimum.

\begin{figure*}[t]
\centering
\includegraphics[width=1.0\linewidth]{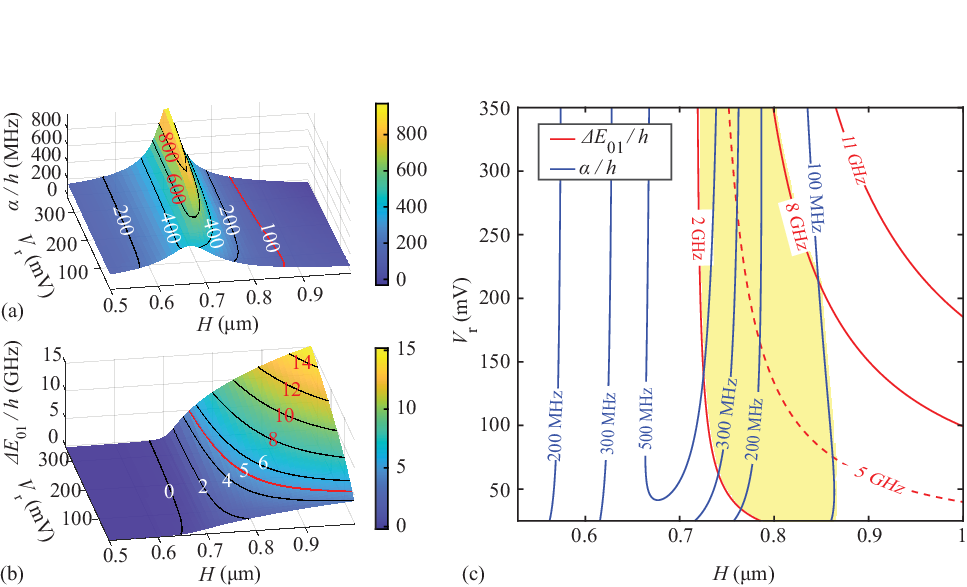}
\caption{(a) Calculated anharmonicity $\alpha=\Delta E_{02}-2\Delta E_{01}$ as a function of $V_r$ and $H$ for the configuration with $R_r=1.5$~$\mu$m, $R_s=0.5$~$\mu$m, and $B_0=-20$~mT. (b) Qubit transition frequency $\Delta E_{01}/h$ as a function of $V_r$ and $H$ for the same configuration. (c) Contour plots of $\alpha$ and $\Delta E_{01}$ illustrating the optimal parameter region, highlighted in yellow.}
\label{Fig10}
\end{figure*}

As the topology of the ground-state wavefunction changes during this process, the anharmonicity $\alpha$ exhibits a pronounced peak. Fig.~\ref{Fig10}(a) shows the calculated $\alpha$ as a function of $V_r$ and $H$ for the same configuration with $R_r = 1.5$~$\mu$m, $R_s = 0.5$~$\mu$m, and $B_0 = -20$~mT. The anharmonicity peaks at $H \approx 0.7$~$\mu$m, and as $V_r$ increases, values of $\alpha / h$ exceeding 0.8~GHz can be achieved. Such a large and tunable anharmonicity is highly desirable, as it allows flexible control of the spectral separation between the qubit transition and higher excited states, beneficial for optimizing gate fidelity and reducing crosstalk in multi-qubit systems. Unlike typical superconducting qubits, whose anharmonicity is negative \cite{Blais2021}, our system exhibits a positive $\alpha$. This positive anharmonicity could help suppress leakage to higher levels: because the microwave drive pulse typically has a frequency tail that falls off toward higher frequencies \cite{Ithier2005}, excitations to higher states are much less likely. As a result, stronger microwave drives can be applied for faster and more selective qubit control.

Nonetheless, as shown in Fig.~\ref{Fig10}(b), the transition frequency $\Delta E_{01}/h$ relevant for qubit operation also varies with $V_r$ and $H$. Ideally, one requires not only a large $\alpha$ for unambiguous state selectivity but also that $\Delta E_{01}/h$ remains within the GHz range for convenient on-chip microwave control. To identify an optimal parameter regime that balances both $\alpha$ and $\Delta E_{01}$, we show in Fig.~\ref{Fig10}(c) contour plots of $\alpha/h$ and $\Delta E_{01}/h$ in the $V_r$–$H$ parameter space. Remarkably, a favorable operating region can be identified, where $\Delta E_{01}/h$ lies in the few-GHz range and $\alpha$ consistently exceeds 100~MHz, comparable to that of typical superconducting qubits~\cite{Kjaergaard2020}.

\section{Electron–resonator coupling and inter-qubit connectivity}\label{Sec:V}
Having established the eigenenergy structure of the electron bound on the SNe microparticle, we now turn to its coupling to the on-chip microwave resonator, which enables coherent control and dispersive readout of the qubit. This coupling arises from the electric-dipole interaction between the electron’s lateral motion and the in-plane electric field of the fundamental differential mode of the $\lambda/4$ CPW resonator. In this mode, opposite-phase voltages on the CPW signal pins generate a strong oscillating field at the open end of the resonator and a node at the shorted end; placing the qubit near the open end therefore maximizes the dipole interaction. The interaction Hamiltonian is given by $ \hat H_{\mathrm{int}} = -\hat{\bm d}\cdot\hat{\bm E} $, where $ \hat{\bm d} = -e\hat{\bm r} $ is the electric-dipole operator and $ \hat{\bm E} $ is the field operator of the resonator mode.

Since the qubit is encoded in the two lowest azimuthal states $\ket{g}\equiv\ket{\psi_{00}}$ and $\ket{e}\equiv\ket{\psi_{01}}$, it is more convenient to express the dipole operator in the spherical basis, where angular-momentum selection rules naturally appear. In this representation, the scalar product can be written as $\hat{\bm{d}}\cdot\hat{\bm{E}}=\sum_{m=-1}^{1}\hat d_{m}\hat E_{-m}$, with dipole components defined in terms of the Cartesian ones as $\hat d_{\pm1}=\mp(\hat d_x\pm i\hat d_y)/\sqrt{2}$ and $\hat d_0=\hat d_z$, and the field components $\hat E_m$ defined analogously. For an electron confined to a spherical surface of radius $ R_s $, the dipole components reduce to $ \hat d_m = -eR_s\sqrt{4\pi/3}Y_{1m}(\theta,\phi) $. Projecting these operators onto the ${\ket{g},\ket{e}}$ subspace yields $\hat d_{+1}=\langle e|\hat d_{+1}|g\rangle \hat\sigma^+ $ and $ \hat d_{-1}=\langle g|\hat d_{-1}|e\rangle \hat\sigma^- $, where $\hat\sigma^+ =\ket{e}\bra{g}$ and $\hat\sigma^-=\ket{g}\bra{e}$ are the qubit ladder operators.

\begin{figure}[t]
\centering
\includegraphics[width=1.0\linewidth]{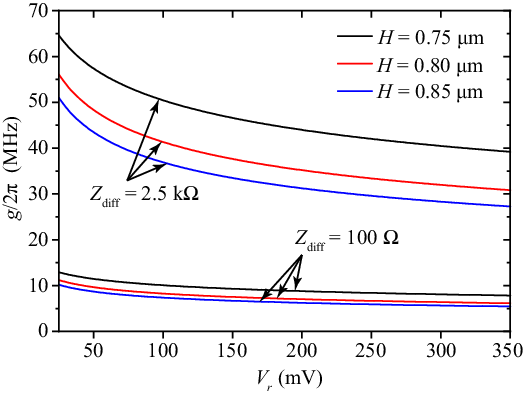}
\caption{Calculated coupling strength $g$ between the eNe qubit and the $\lambda/4$ microwave resonator versus the applied resonator bias voltage $V_r$, shown for various separation distances $H$ and differential impedances $Z_{\mathrm{diff}}$.}
\label{Fig11}
\end{figure}

The electric field of the differential mode oscillates primarily along the in-plane $x$ direction, as shown in Fig.~\ref{Fig7}. This field can be written as $\hat E_x = E_{\mathrm{zpf}}(\hat a + \hat a^\dagger)$, where $\hat a$ and $\hat a^\dagger$ are the photon annihilation and creation operators of the resonator mode, and $E_{\mathrm{zpf}}$ is the zero-point field amplitude. The latter is given by $E_{\mathrm{zpf}} = \mathcal{E}_V V_{\mathrm{zpf}}$, where $\mathcal{E}_V$ is the field strength at the qubit location per volt of differential excitation on the resonator, which can be obtained from electrostatic simulations (see Appendix~\ref{App-E}), and $V_{\mathrm{zpf}} = \omega_r\sqrt{\hbar Z_{\mathrm{diff}}/2}$ is the zero-point voltage of a resonator with resonance frequency $\omega_r$ and differential impedance $Z_{\mathrm{diff}}$~\cite{Koolstra2025highimpedance}. Using the spherical decomposition of the field and noting that $\hat E_y\simeq\hat E_z\simeq0$, we obtain $\hat E_{\pm1}=\mp\hat E_x/\sqrt{2}=\mp E_{\mathrm{zpf}}(\hat a + \hat a^\dagger)/\sqrt{2}$ and $\hat E_0 = 0$.

Substituting the spherical components of $\hat{\bm d}$ and $\hat{\bm E}$ into $\hat H_{\mathrm{int}}$ and retaining only the energy-conserving terms (rotating-wave approximation) yields the standard Jaynes–Cummings form $ \hat H_{\mathrm{int}} \simeq \hbar g(\hat a\hat\sigma^+ + \hat a^\dagger\hat\sigma^-) $, with the coupling strength:
\begin{equation}
\frac{g}{2\pi} = \frac{\omega_r}{2\pi\hbar}\sqrt{\frac{\hbar Z_{\mathrm{diff}}}{2}}\mathcal{E}_V\bigl|\langle e|\hat d_{+1}|g\rangle\bigr|,
\end{equation}
where the dipole matrix element $\langle e|\hat d_{+1}|g\rangle$ can be obtained numerically from the wavefunctions calculated in Sec.~\ref{Sec:IV}. Using a typical differential impedance $Z_{\mathrm{diff}} \simeq 100~\Omega$, corresponding to an odd-mode characteristic impedance of 50~$\Omega$ per CPW signal line, and resonance frequency $\omega_r/2\pi=5$~GHz, the resulting coupling $g/2\pi$ as a function of the resonator bias voltage $V_r$ for various separations $H$ is plotted in Fig.~\ref{Fig11}. The coupling strength increases with decreasing $V_r$ and $H$, and remains above 5~MHz throughout the explored parameter space, ensuring strong coupling given observed eNe linewidths of $\lambda/2\pi\sim0.1$~MHz~\cite{Zhou2024}. Furthermore, high-impedance resonators based on TiN with $Z_{\mathrm{diff}}\simeq2.5$~k$\Omega$ have recently been demonstrated~\cite{koolstra2025strongcouplingmicrowavephoton}, which could boost $g/2\pi$ into the tens-of-MHz range as shown in Fig.~\ref{Fig11}, on par with that of superconducting qubits.

\begin{figure}[h]
\centering
\includegraphics[width=1.0\linewidth]{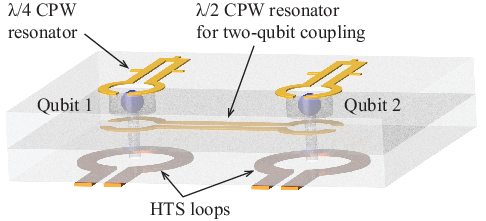}
\caption{Schematic showing how two eNe qubits could be coupled using a $\lambda/2$ CPW resonator.}
\label{Fig12}
\end{figure}

To implement entanglement and universal two-qubit gate operations, it is necessary to coherently couple multiple eNe qubits. This can be achieved using the same resonator-mediated architecture widely adopted in superconducting circuit-QED systems. In close analogy to those platforms, two eNe qubits can be coupled to a shared microwave resonator mode, which serves as a quantum bus enabling coherent interactions through virtual-photon exchange~\cite{Majer2007}. For this purpose, one may employ a $\lambda/2$ differential CPW resonator, as shown schematically in Fig.~\ref{Fig12}. The $\lambda/2$ resonator supports voltage antinodes at both ends, allowing one qubit to be placed at each antinode. Let the two eNe qubits placed at these antinodes have transition frequencies $\omega_{q,1}$ and $\omega_{q,2}$ and couple to the resonator mode with strengths $g_1$ and $g_2$, respectively. When both qubits are detuned from the resonator frequency $\omega_r$ such that $|\omega_{q,k}-\omega_r|\gg g_k$, the system lies in the dispersive regime, where the resonator remains in its vacuum state and mediates only virtual-photon processes~\cite{Majer2007}. The dynamics are then governed by the Hamiltonian $ \hat H = \hbar\omega_r\hat a^\dagger\hat a + \sum_{k=1}^2 (\hbar\omega_{q,k}/2)\hat\sigma_z^{(k)} + \hbar g_k(\hat a\hat\sigma_+^{(k)} + \hat a^\dagger\hat\sigma_-^{(k)}) $. Eliminating the resonator mode via a second-order Schrieffer–Wolff transformation yields an effective exchange interaction between the two qubits~\cite{Blais2021}: $ \hat H_{\mathrm{eff}} = \hbar J(\hat\sigma_+^{(1)}\hat\sigma_-^{(2)} + \hat\sigma_-^{(1)}\hat\sigma_+^{(2)}) $, where $ J = g_1 g_2 / \Delta $ and $\Delta = \omega_q - \omega_r$ is the qubit–resonator detuning (assuming $\omega_{q,1}\simeq\omega_{q,2}=\omega_q$). For representative parameters, i.e., $g_k/2\pi$ in the 10–30~MHz range and detunings $|\Delta|/2\pi$ of 50–150~MHz, the resulting two-qubit coupling strength is $J/2\pi\sim2$–6~MHz. Such coupling strengths comfortably exceed decoherence rates, suggesting that high-fidelity two-qubit entangling operations should be achievable within this resonator-mediated architecture.

\section{Discussion}\label{Sec:VI}
The results presented here establish that electrons bound to on-chip levitated SNe microparticles provide a robust and highly tunable platform for qubit implementation. In particular, we have shown that the lateral motional states of a surface-bound electron can be engineered into a microwave qubit with large positive anharmonicity, strong and controllable coupling to on-chip superconducting resonators, and wide-range electrical tunability. Moreover, the same resonator infrastructure naturally enables coherent coupling between neighboring eNe qubits, providing a scalable route toward multi-qubit connectivity. Beyond the motional-state qubit demonstrated here, this architecture also accommodates additional qubit modalities and hybrid operation.

A particularly compelling opportunity is the use of the electron’s spin degree of freedom, whose coherence is normally limited by magnetic noise from surrounding nuclear spins. SNe is highly advantageous in this respect: although natural neon contains small fractions of isotopes with nonzero nuclear spin, isotopically purified $^{20}\mathrm{Ne}$ is commercially available and provides an effectively nuclear-spin–free host. Indeed, theoretical modeling for an electron bound to a purified SNe surface predicts spin coherence times as long as 81~s~\cite{chen2022electron}, placing the system among the longest-lived solid-state spins~\cite{BarGill2013,Tyryshkin2011}. The levitated SNe architecture further enhances this potential by isolating the electron from substrate-induced charge fluctuations, magnetic noise, and parasitic paramagnetic impurities, allowing the intrinsic spin coherence to be fully realized in practice. Spin manipulation requires only a modest static magnetic field ($B\sim0.1$–$0.3~\mathrm{T}$) to define the Zeeman splitting, fully compatible with NbTiN/NbN resonators that retain high $Q$ in this field range. Although magnetic dipole coupling is intrinsically weaker than the electric dipole interaction used for the charge motional qubit, making coherent coupling of two spin qubits via single virtual microwave photons challenging in standard CPW architectures, suitably engineered resonators such as lumped element structures with tightly confined magnetic fields could still enable high-fidelity dispersive readout and cavity assisted control of individual spins. In this framework, eNe spin qubits can serve as long-lived quantum memories coexisting on the same chip with fast, high-anharmonicity eNe charge qubits, enabling hybrid architectures where rapid charge-based gates interface with long-coherence spin storage, a key capability for modular quantum processors and future quantum communication networks.

More broadly, the underlying principles of magnetic levitation and cavity integration can be extended to other quantum fluid and solid (QFS) systems. In particular, arrays of magnetically levitated He~II microdroplets can be realized using the same superconducting-loop architecture and loading protocols developed for SNe particles. These droplets provide ultraclean environments ideal for hosting dopant ions such as $\mathrm{Er^{3+}}$, which offer telecom-band optical transitions and millisecond spin coherence times~\cite{Bttger2009,Probst2015,Ourari2023}. In He II, each embedded positive ion is known to form a nanometer-scale ``snowball'' structure that generates a well-defined local crystal field~\cite{Atkins1959,Johnson1972}, producing robust level splittings of $\mathrm{Er^{3+}}$ ions that further facilitate optical or spin-qubit level selection~\cite{Bottger2009, Ourari2023}. Embedding $\mathrm{Er^{3+}}$ ions in levitated He~II droplets yields naturally isolated quantum emitters with minimal dielectric loss, while the droplets themselves could act as high-finesse whispering-gallery-mode resonators that enhance light–matter interaction~\cite{Brown2023}. Such droplet nodes could operate as quantum repeaters for long-distance entanglement distribution, bridging microwave and optical domains through hybrid spin–photon interfaces~\cite{Uysal2024}. The ability to co-fabricate arrays of SNe and He~II carriers on the same chip further suggests a unified quantum platform where motional or spin-based eNe qubits interface coherently with $\mathrm{Er^{3+}}$ optical nodes, enabling modular and reconfigurable quantum networks. Altogether, these capabilities underscore the broader significance of the levitated QFS architecture as a versatile foundation for scalable quantum processing, distributed communication, and hybrid quantum technologies.

\begin{acknowledgments}
The authors acknowledge the support from the Gordon and Betty Moore Foundation through Grant DOI 10.37807/gbmf11567, the National Science Foundation (NSF) under Award No. OSI-2426768, and the FSU-QuEST Award through the Florida State University Quantum Initiative Program. The authors also thank Xianjing Zhou, Lukasz Dusanowski, Tongcang Li, Jack Harris, Peng Xiong, Irinel Chiorescu, Lance Cooley, and William Oates for valuable discussions. The work was conducted at the National High Magnetic Field Laboratory at Florida State University, which is supported by the National Science Foundation Cooperative Agreement No. DMR-2128556 and the state of Florida.
\end{acknowledgments}

\appendix
\section{Magnetic field calculations}\label{App-A}
The magnetic field $\mathbf{B}_L(\mathbf{r})$ generated at position $\mathbf{r}$ by an ideal zero-thickness current loop in three-dimensional space can be calculated using the Biot-Savart law~\cite{jackson1999classical}:
\begin{equation}
\mathbf{B}_L(\mathbf{r}) = \frac{\mu_0 I}{4\pi}
\int \frac{d\mathbf{l}\times(\mathbf{r}-\mathbf{l})}{|\mathbf{r}-\mathbf{l}|^{3}},
\label{eq:biot-savart}
\end{equation}
where $d\mathbf{l}$ is the elemental length vector along the loop.

To model the magnetic field produced by a circular current-carrying disk with inner radius $R_0$ and width $W$, we approximate the disk using $n$ uniformly spaced concentric loops. Each loop carries an equal fraction of the total current, $I/n$. Owing to axial symmetry, the magnetic field distribution can be evaluated in the two-dimensional $xz$-plane, with the $z$-axis aligned with the disk axis and the origin located at the geometric center of the disk. The contributions to the magnetic field components at a point $(x,z)$ from the discretized disk are then given by:
\begin{align}
B_x(x,z) &= \sum_{i=1}^{n}
\frac{\mu_0 I}{4\pi n}
\int_{0}^{2\pi}
\frac{R_i z \cos\phi}{D_i^{3}}\, d\phi, \label{eq:Bx} \\
B_z(x,z) &= \sum_{i=1}^{n}
\frac{\mu_0 I}{4\pi n}
\int_{0}^{2\pi}
\frac{R_i^{2} - R_i x \cos\phi}{D_i^{3}}\, d\phi,
\label{eq:Bz}
\end{align}
where
\begin{align}
D_i &= \sqrt{[x - R_i\cos\phi]^{2} + [R_i\sin\phi]^{2} + z^{2}}, \\
R_i &= R_0 + \left(i - \frac{1}{2}\right)\frac{W}{n}.
\end{align}
In our simulations, the computational domain is discretized on a square grid with spatial resolution $\Delta x=\Delta z=0.1~\mu\mathrm{m}$, and the disk is represented using $n = 30$ concentric loops.

\section{WKB Calculation of the Electron Tunneling Time}\label{App-B}
When no external field is applied, the electron’s wavefunction spreads over the entire SNe surface. Applying a positive DC voltage to the resonator generates an electric field
throughout the surrounding space, which pulls the electron toward the north pole of the SNe particle. Owing to the axial symmetry of the system, the electric field at the pole is directed essentially normal to the surface; its magnitude at that point is denoted by $E_r$. This normal field tilts the vertical image-binding potential and lowers the surface barrier that confines the electron. The effective vertical potential governing escape is therefore:
\begin{equation}\label{U_p}
U_{\perp}(z)
=
-\frac{\varepsilon - \varepsilon_0}{\varepsilon + \varepsilon_0}
\frac{e^2}{16\pi\varepsilon_0 z}
- e E_r z,
\qquad z>0 ,
\end{equation}
where $z$ is the height above the surface, $\varepsilon_0$ is the vacuum permittivity, and $\varepsilon = 1.244\,\varepsilon_0$ is the dielectric constant of SNe. For sufficiently large $E_r$, the tilted potential becomes low and narrow enough such that the electron can escape by quantum tunneling through the energy barrier. The corresponding tunneling lifetime can be evaluated using the WKB approximation.

Following Ref.~\cite{Cole1971}, we regularize the polarization term in $U_{\perp}(z)$ by truncating it at $z=b=2.3~\text{\AA}$ and extending it to the origin as a constant value $-\frac{\varepsilon - \varepsilon_0}{\varepsilon + \varepsilon_0}\frac{e^2}{16\pi\varepsilon_0 b}$. Let $\epsilon_1$ denote the ground-state energy (including the first-order Stark shift). The classical turning points $z_1$ and $z_2$ are defined implicitly by:
\[
    U_{\perp}(z_1)=U_{\perp}(z_2)=\epsilon_1 .
\]
Within the WKB approximation, the tunneling lifetime is given by~\cite{SakuraiQM}:
\begin{equation}
    \tau_{\mathrm{WKB}}
    = T_{\mathrm{el}}
      \exp\!\left[
      \frac{2}{\hbar}
      \int_{z_1}^{z_2}
      \sqrt{2m_e\left(U_{\perp}(z)-\epsilon_1\right)}\,dz
      \right],
    \label{eq:AppC_tau}
\end{equation}
where $m_e$ is the electron mass. The prefactor $T_{\mathrm{el}}$ is the classical oscillation period in the bound region:
\begin{equation}
    T_{\mathrm{el}}
    = 2
      \int_{0}^{z_1}
      \sqrt{\frac{m_e}{2\left(\epsilon_1-U_{\perp}(z)\right)}}\,dz ,
    \label{eq:AppC_Tel}
\end{equation}
which is evaluated numerically using the expression for $U_{\perp}(z)$ given in Eq.~(\ref{U_p}). The resulting tunneling lifetimes $\tau_{\mathrm{WKB}}$, plotted in Fig.~6(b), decrease rapidly from the order of seconds to sub-millisecond values as $E_r$ crosses a threshold region around $0.4~\mathrm{V}/\mu\mathrm{m}$.

\section{Electrostatic Potential Generated by the Biased Resonator}\label{App-C}
To obtain an analytic expression for the lateral confinement potential $U_{\parallel}(\theta)$ used in Eq.~(2), we model the open end of the $\lambda/4$ CPW resonator as a thin circular ring of radius $R_r$ held at a DC voltage $V_r$ and located at a distance $H$ above the center of the SNe particle. In cylindrical coordinates $(\rho,z)$ with the ring lying in the plane $z = H$, the potential energy of an electron at position $(\rho,z)$ generated by this ring can be written in closed form using complete elliptic integrals~\cite{jackson1999classical}:
\begin{equation}
    U(\rho,z)
    = 2 e V_r\,
      \frac{K(k)}{
      K_{\mathrm{eff}}\,
      \sqrt{(R_r + \rho)^2 + (z - H)^2}},
    \label{eq:AppB_Urho}
\end{equation}
where $K(k)$ is the complete elliptic integral of the first kind, and the elliptic modulus is:
\begin{equation}
    k^2 =
    \frac{4 R_r \rho}{
      (R_r + \rho)^2 + (z-H)^2 }.
\end{equation}
The normalization factor $K_{\mathrm{eff}}$ ensures that the potential scale is consistent with the applied voltage $V_r$.  In our numerical
implementation, $K_{\mathrm{eff}}$ is chosen such that
\[
    U(\rho = R_{r} - a_{r},\, z = H) = -eV_{r},
\]
where $a_{r}$ represents the effective radial half-width of the resonator’s center pins. This condition anchors the potential at the pin edge.

To obtain the angular confinement potential on the SNe particle, we restrict the electron to the sphere of radius $R_s$ through the substitutions:
\[
    \rho = R_s \sin\theta, \qquad
    z    = R_s \cos\theta,
\]
which yields the expression $U_{\parallel}(\theta)$ used in Eq.~(\ref{eq:Hamiltonian}) of the main text.

\section{Calculation of the Electron Lateral Eigenstates}~\label{App-D}
The lateral eigenstates of the surface-bound electron can be obtained by solving the Hamiltonian in Eq.~(\ref{eq:Hamiltonian}) of the main text. Owing to the axial symmetry of the system, the azimuthal quantum number $m$ is conserved, and the wavefunction can be written as $\Psi_{m}(\theta,\phi)=\frac{1}{\sqrt{2\pi}}\psi_{m}(\theta)e^{i m\phi}$. Our numerical procedure then proceeds in two steps. First, we construct a complete set of trial eigenstates by diagonalizing a simplified Hamiltonian whose matrix elements are analytic in the spherical-harmonics basis. These trial states are then refined by imaginary-time evolution under the full Hamiltonian on a real-space $\theta$ grid, with orthogonality enforced throughout so that the entire eigenstate spectrum is faithfully recovered.

To construct the simplified Hamiltonian, we approximate the lateral potential $U_{\parallel}(\theta)$ near the north pole of the SNe particle as:
\begin{equation}
U_{\parallel}(\theta)
\simeq
U_{\parallel}(0) + eE_r R_s\bigl(1 - \cos\theta\bigr),
\end{equation}
where $E_r=[\partial U(0,z)/\partial z]_{z=R_s}$ is the electric-field strength at the north pole, and $U(\rho,z)$ is given analytically in Appendix~\ref{App-C}. The simplified Hamiltonian then takes the form:
\begin{align}
\hat{H}_{\parallel}^{(s)}
&=
-\frac{\hbar^{2}}{2 m_e R_s^{2}}\,\nabla^{2}_{\theta\phi}
+ \frac{e^{2} B^{2} R_s^{2}}{8 m_e}\,\sin^{2}\theta
+ \frac{eB}{2m_e}\,m\hbar
\nonumber\\[4pt]
&\quad
+ U_{\parallel}(0)
+ eE_r R_s\bigl(1 - \cos\theta\bigr).
\end{align}
All operators appearing in this simplified Hamiltonian, such as the angular derivatives in $\nabla^{2}_{\theta\phi}$, $\cos\theta$, and $\sin^{2}\theta$, have
closed-form analytic matrix elements in the spherical-harmonics basis $Y_{l,m}(\theta,\phi)$~\cite{schmidt2015supersymmetry,PhysRevB.46.9501}. For each $m$, diagonalizing $\hat{H}_{\parallel}^{(s)}$ in this basis yields a complete set of trial eigenfunctions of the form: 
\begin{equation}
\psi^{(0)}_{n m}(\theta)e^{i m \phi}
=
\sum_{l = |m|}^{N_{max}}
c^{(n m)}_{l}\,Y_{l,m}(\theta,\phi),
\end{equation}
where the coefficients $c^{(n m)}_{l}$ are the components of the $n$-th eigenvector. A finite cutoff $N_{max}$ is required to truncate the spherical-harmonics basis to
a manageable matrix size while still resolving the narrow angular localization of the eigenstates near the north pole.  We set $N_{max}=800$, which provides an angular
resolution of about $4\times 10^{-3}\,\mathrm{rad}$, sufficient to capture the spatial extent of the confined states. These trial eigenfunctions $\{\psi^{(0)}_{n m}\}$ serve as the initial states for refining the eigenfunctions of the full Hamiltonian.

In the second step, the full Hamiltonian $\hat{H}_{\parallel}$ shown in Eq.~\eqref{eq:Hamiltonian} is solved by imaginary-time evolution. For numerical integration, we work with the dimensionless Hamiltonian $\hat{\mathcal{H}}=\hat{H}_{\parallel}/E_{0}$, where $E_{0}=\hbar^{2}/(2m_{e}R_{s}^{2})$ is the natural energy scale associated with a free electron on the SNe particle surface. Introducing the dimensionless imaginary-time variable $\tau = i\,(E_{0}/\hbar)\,t$, the evolution equation becomes:
\begin{equation}
\frac{\partial\psi(\theta,\tau)}{\partial\tau}
=
-\hat{\mathcal{H}}\,\psi(\theta,\tau).
\end{equation}
Each state is initialized as $\psi_{n m}(\theta,\tau=0)=\psi^{(0)}_{n m}(\theta)$, and the equation is integrated iteratively using a forward–Euler step:
\begin{equation}
\psi(\theta,\tau+\delta\tau)
=
\psi(\theta,\tau)
-\delta\tau\,\hat{\mathcal{H}}\psi(\theta,\tau),
\end{equation}
with dimensionless time step $\delta\tau = 10^{-6}$. Because the wavefunction norm is not conserved under imaginary-time propagation, the state is renormalized after every iteration by multipling a factor $[\int_{0}^{\pi} |\psi(\theta,\tau)|^{2}\sin\theta\,d\theta]^{-1/2}$. Angular derivatives in $\hat{H}_{\parallel}$ are evaluated using second-order finite differences on a uniform $\theta$ grid with a resolution $\Delta\theta = 3.1\times 10^{-3}\,\mathrm{rad}$.

At every iteration, orthogonality among all states with the same $m$ is enforced using:
\begin{equation}
\langle\psi_{j m}|\psi_{k m}\rangle
=
\int_{0}^{\pi}
\psi_{j m}^{*}(\theta)\,\psi_{k m}(\theta)\,
\sin\theta\,d\theta,
\end{equation}
together with the projection:
\begin{equation}
\psi_{k m}(\tau)
\leftarrow
\psi_{k m}(\tau)
-
\sum_{j<k}
\langle\psi_{j m}|\psi_{k m}(\tau)\rangle\,\psi_{j m}(\tau).
\end{equation}
The iterations continue until both the wavefunctions and their energies $E_{n m}=\langle\psi_{n m}|\hat{H}_{\parallel}|\psi_{n m}\rangle$ have converged. These converged eigenstates and eigenenergies form the basis for the qubit transition frequencies, anharmonicity, and dipole matrix elements presented in the main text.

\section{Differential-Mode Electric Field at the North Pole of the SNe Particle}~\label{App-E}
The electric dipole coupling between the qubit and the resonator is determined by the microwave electric field generated at the SNe particle by the resonator’s differential
mode. This AC field is entirely distinct from the static field produced by the DC bias applied simultaneously to both resonator pins. The DC bias generates a common-mode
electric field that shapes the lateral confining potential $U_{\parallel}(\theta)$, whereas the differential-mode field provides the oscillating drive that enters the
interaction Hamiltonian $\hat H_{\mathrm{int}}$ and ultimately sets the strength of the electron–resonator coupling $g$.

\begin{figure}[t]
\centering
\includegraphics[width=1.0\linewidth]{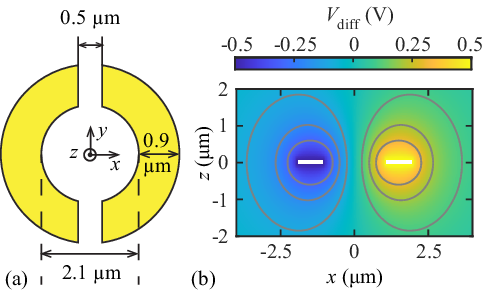}
\caption{(a) Geometry and dimensions of the resonator center pins used in the MATLAB simulation of the differential mode. (b) Differential-mode potential $V_{\mathrm{diff}}$ in the $x$–$z$ plane obtained from the simulation.}
\label{Fig13}
\end{figure}

To quantify this coupling, we require the electric-field amplitude at the qubit location per volt of differential excitation, defined as $\mathcal{E}_V$.  This quantity is
obtained from electrostatic simulations performed in MATLAB, in which the two resonator center pins are driven in differential mode.  The geometry and dimensions of the pins
used in the simulation are shown in Fig.~\ref{Fig13}(a).  In the calculation, the left and right pins are biased at $-0.5~\mathrm{V}$ and $+0.5~\mathrm{V}$, respectively, so that the applied differential voltage is exactly $1~\mathrm{V}$, while all ground planes and surrounding conductors are held at $0~\mathrm{V}$.  Solving Laplace’s equation under these boundary conditions yields the differential-mode potential $V_{\mathrm{diff}}(\mathbf{r})$ shown in Fig.~\ref{Fig13}(b).  The corresponding electric field is obtained from $\mathbf{E}_{\mathrm{diff}}(\mathbf{r}) = -\nabla V_{\mathrm{diff}}(\mathbf{r})$, whose in-plane $x$ component dominates near the qubit.  Evaluating this field at the SNe particle’s north pole $(\rho = 0, z = R_s)$ gives the local differential-mode amplitude $E_{\mathrm{diff}}(0,R_s)$, which by construction renders $\mathcal{E}_V =
E_{\mathrm{diff}}(0,R_s)$ because the applied differential excitation is $1~\mathrm{V}$.

\bibliography{Ref-LeviQ}

\end{document}